\documentclass{article}

\usepackage{arxiv}

\usepackage[utf8]{inputenc} %
\usepackage[T1]{fontenc}    %
\usepackage{hyperref}       %
\usepackage{url}            %
\usepackage{booktabs,longtable}       %
\usepackage{siunitx}
\usepackage{amsfonts}       %
\usepackage{nicefrac}       %
\usepackage{microtype}      %
\usepackage{lipsum}
\usepackage{xcolor}
\usepackage{float}

\usepackage{mathtools}
\usepackage{float}
\usepackage[style=nature,backend=biber,doi=false,url=false]{biblatex}
\usepackage[version=3]{mhchem}

\title{{A General Topological Network Criterion for Exploring the Structure of Icy Nanoribbons and Monolayers}}

\author{
	\href{https://orcid.org/0000-0001-8706-2383}{\includegraphics[scale=0.06]{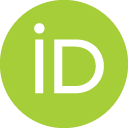}\hspace{1mm}Amrita Goswami} \\
	Department of Chemical Engineering\\
	Indian Institute of Technology Kanpur\\
	\texttt{amritag@iitk.ac.in} \\
	\And
	\href{https://orcid.org/0000-0001-8056-2115}{\includegraphics[scale=0.06]{orcid.png}\hspace{1mm}Jayant K. Singh}\thanks{\textbf{Corresponding Author}} \\
	Department of Chemical Engineering\\
	Indian Institute of Technology Kanpur\\
	\texttt{jayantks@iitk.ac.in} \\
}

\begin{document}
\maketitle

\begin{abstract}
	We develop intuitive metrics for quantifying complex nucleating
	systems under confinement. These are shown to arise naturally from the analysis of the topological ring network, and are amenable for use as order parameters for such systems. Drawing inspiration from qualitative visual inspection, we introduce a general topological criterion for elucidating the ordered structures of confined water, using a graph theoretic approach. Our criterion is based on primitive rings, and reinterprets the hydrogen-bond-network in terms of these primitives. This approach has no a priori assumptions, except the hydrogen bond definition, and may be used as an exploratory tool for the automated discovery of new ordered phases. We demonstrate the versatility of our criterion by applying it to analyse well-known monolayer ices. Our methodology is then extended to identify the building blocks of one-dimensional \(n\)-sided prismatic nanoribbon ices.
\end{abstract}


\section{Introduction}\label{introduction}
Confinement is known to cause deviations from bulk system properties
\cite{Koga2001,Kalra2003,Walther2013,Takaiwa2007}, imparting
unique complexities to the structure determination of icy confined
water. The structure and phase behaviour of quintessential confined
systems, like that of two-dimensional monolayer and one-dimensional
nanoribbons are of enormous
theoretical\cite{Zhao2014,Takaiwa2007,Bai2010} and experimental
interest\cite{Algara-Siller2015,Agrawal2016}. However, well-defined
simulation methodologies and methods of analysis for bulk systems like
the mean squared displacement (MSD), often used as a broad indicator
of phase transitions, may be unreliable for constrained ice systems.
The reasons are two-fold: confined water often exhibits continuous
freezing transitions \cite{Mochizuki2015, Zhu2015, Zhu2016}, and the
mobility of water layers close to the confining sheet is lower than
that of intermediate layers \cite{Hirunsit2007, Mosaddeghi2012}.
Parameters engineered to characterize phase transitions in simpler
Lennard-Jones systems, inspired by the MSD, are not
universally valid. For example, the Lindemann parameter, which has
been used for melting studies\cite{Gilvarry1966, Zheng1998}, fails for
confined systems with attractive pores due to instabilities in the MSD
\cite{Das2013}.  Structure determination techniques, which are
accurate for bulk water, are often intractable for nucleating water
systems under confinement, as the hydrogen-bonding network (HBN) is
significantly influenced by the confining wall. Compared to bulk
water, water under nanoscale confinement shows considerable
polymorphic diversity \cite{Zhao2014a}.

Visual inspection of the HBN, paired with the angle distribution, has
been used in the literature for the analysis of confined water
systems. Such techniques are often used for qualitative analyses
\cite{Zhu2015,Zhu2016,Yang2017,Zhao2014,Chen2016} but are unable to
describe the structures quantitively. Voronoi tessellation
\cite{Zhu2017} has been previously used to quantify the structures of
two-dimensional confined ice systems. The bond orientational parameter
has also been used for distinguishing the degree of square and
hexagonal order in ice \cite{Dix2018}. However, the bond orientational
parameter has implicit a priori assumptions of the ice structures
formed, and cannot be used as an exploratory tool. Moreover, none of
these parameters has been successfully applied to one-dimensional
nanoribbon ices, although axial and angular order parameters for
distinguishing between square and pentagonal ice nanotubes have been
used \cite{Koga2001}. There is no single straightforward
classification technique for confined ordered ice-like water that minimizes
human involvement and assumptions, and emphasizes automated,
reproducible, quantifiable structural elucidation.

Determining the connectivity of an ordered phase by using the
primitive rings formed is a well-established technique of
identification
\cite{Yuan2001,Chihaia2005,Salzmann2011,Goetzke1991,Zhang2017,Prerna2019}.
Ic (cubic ice) and Ih (hexagonal ice) are ice phases, which are
similar enough that their structural differences are not trivially
discernible by casual inspection. A robust and intuitive
classification of the building-blocks of Ic and Ih, based on primitive
ring connectivity, has been formulated \cite{HajiAkbari2015} and used
for homogenous and heterogenous bulk nucleation
\cite{Sosso2016,Saito2018,Sosso2016a}. However, these topological
network connectivity analyses, like other popular analyses including
spherical harmonics based methods, suffer from a common shortcoming.
All such classification techniques use a fixed-distance cutoff, after
which further reductions or criteria are applied. These techniques
work on the assumption that distance-based information is nearly
equivalent to the HBN information. This implicit assumption is known
to be valid for bulk systems of deeply supercooled water
\cite{Sciortino1989}. Thus, for such systems, the distance-based
cutoff, corresponding to the first nearest-neighbor shell of the
oxygen atoms (first minimum of the O-O radial distribution function)
is a close approximation of the HBN, with negligible overcounting
\cite{HajiAkbari2015}.

The scenario is more complex for confined systems. Confined systems
show a variety of ordered phases at temperatures much higher than the
freezing point of the bulk liquid. Experimental results have predicted
substantial temperature elevations of the freezing transition for
water constrained within isolated carbon nanotubes, compared to the
bulk phase \cite{Agrawal2016}. Simulations have also shown that
certain confined systems exhibit a freezing point as high as \(390 \ K\)
\cite{Pugliese2017}. In particular, at high pressures or under strong
confinement, we show that a fixed distance-cutoff creates connections
between molecules which are not actually part of the HBN. A naive
application of a topological network criterion based on a fixed
distance cutoff, thus, necessarily fails by creating extraneous
connections.

We summarize the main practical challenges of confined ice structure
determination as follows:

\begin{enumerate}
	\item Confined water has a high number of possible polymorphs of
	      quasi-one and two-dimensional ice. These manifest even at elevated temperatures
	      compared to the bulk. The ordered structures are
	      highly sensitive to the geometry of the confining surface and
	      hydrophobicity.
	\item A molecule in confined water may not actually form a
	      hydrogen-bond with every one of its nearest neighbours. This is
	      because molecules forced within the first nearest-neighbour shell
	      under the effect of confinement or high pressure may not fulfil
	      the additional strict criterion of hydrogen bond formation. We
	      show that not accounting for the hydrogen bonds can cause
	      significant mis-identification of ice.
\end{enumerate}

In the following sections, we introduce a robust topological network criterion for
quantitatively and qualitatively classifying quasi-one-dimensional and
two-dimensional ice structures, specifically adapted for confined
systems. We demonstrate the generality of our method by using it to
identify structures of well-known one-dimensional nanoribbons and
two-dimensional monolayer ices. We have also formulated a prism
identification scheme for identifying the constituent building blocks
of $n$-gonal one-dimensional ice.  Our
graph-theoretic network criterion leads to intuitive metrics for
quantifying network structures, including the projected area, the
occupied volume, and other similar topological features, which shed
light on simulation results. We prove the efficacy of such analysis
techniques for systems at higher temperature phases and transitions of
monolayer ice within a graphite nanochannel. We show how our
topological network parameter may be used for exploratory studies of
entirely new systems. Our order parameter as described, is based on
purely topological features, and makes no assumptions of the
regularity of the local environment or symmetry factors, instead
relying solely on the description of the hydrogen bonds.

\hypertarget{theory-and-methods}{%
	\section{Theory and Methods}\label{theory-and-methods}}

\hypertarget{topological-network-criterion}{%
	\subsection{Topological Network
		Criterion}\label{topological-network-criterion}}

\begin{figure}[H]
	\centering
	\includegraphics[scale=0.5]{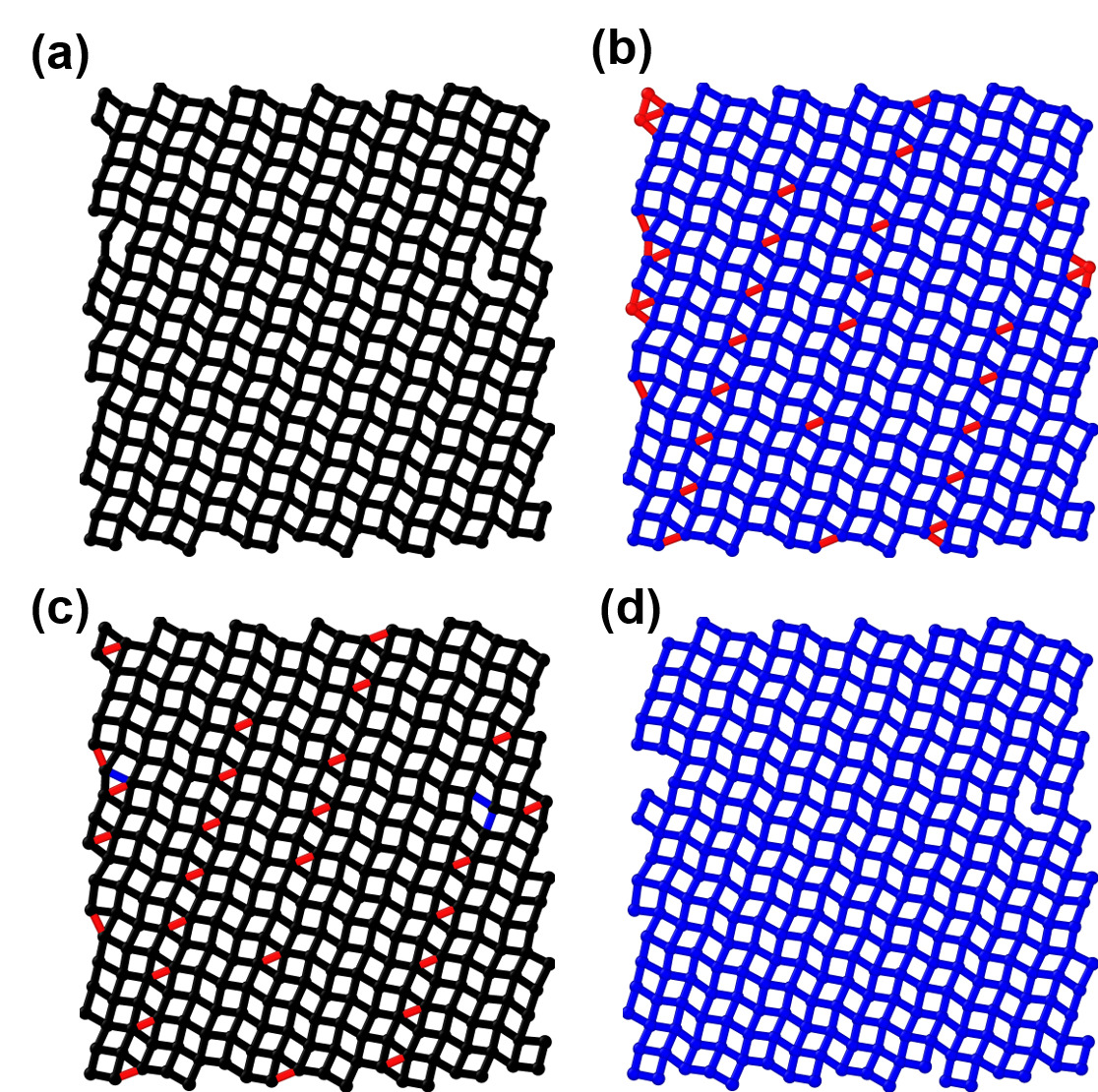}
	\caption{Intermediate stages of our topological network criterion, in the classification of fMSI.(a) Top view of the hydrogen bonding network, determined using a strict geometric criterion. Only oxygen atoms have been shown. Molecules which participate in hydrogen bonds are connected by black bonds as shown. \label{subfig:onlyHBN} (b) 4-membered rings (blue) and 3-membered rings (red), using a fixed-distance cutoff of $3.5$ \si{\angstrom}. Shared edges have been shown in blue. The ring network shown here has been created using primitive rings with a fixed-distance cutoff only. Several extra bonds are introduced. \label{subfig:noAlgo} (c) HBN (black) overlaid over 4-membered rings (blue) and 3-membered rings (red). Extraneous bonds in blue and red are clearly visible. The extra triangular phase (red) is particularly prominent, which has no overlap with the HBN. \label{subfig:extraBonds} (d) 4-membered rings (blue) and 3-membered rings (red) after the application of our topological network criterion, eliminating superfluous bonds. No triangular phase remains. \label{subfig:awesomeAlgo} }
	\label{fig:topoMethod}
\end{figure}

Visual inspection of confined systems (Figure \ref{subfig:onlyHBN}(a)) gravitates towards the identification of polygons. To this end, we have shown that the composition of the HBN, in terms of $n$-membered rings may be used for classification purposes.
Our definition of a ring is similar to that
in the literature \cite{Salzmann2011,HajiAkbari2015}. We define a ring
as the largest polygon that can be formed from oxygen atoms of
hydrogen-bonded molecules, such that it cannot be broken into smaller
constituent rings. This is, in essence, the mathematical description of a primitive ring, using King's shortest path criterion \cite{SHIRLEYV.1967}.

Here, we outline the methodology by applying it to fMSI (flat Monolayer
Square Ice) formed within a graphene nanochannel \cite{Yang2017}. In
previous treatments of bulk ice and supercooled water, primitive ring
analyses have been implemented using a fixed-distance cutoff
\cite{Salzmann2011,HajiAkbari2015}. However, we demonstrate that a naive
application of primitive ring analysis using a fixed-distance cutoff
introduces severe error, necessitating modifications of the methodology
for confined systems. The ring network formed using a fixed-distance cutoff has been depicted in
Figure \ref{subfig:noAlgo}(b). Figure \ref{subfig:extraBonds}(c) shows extraneous bonds, coloured in red
and blue, caused by using primitive rings without accounting for the
HBN, shown here in black. This visual representation shows how the HBN
does not coincide with the ring network formed using a fixed-distance
cutoff.

Our topological network criterion incorporates the information of the
HBN, thus eliminating superfluous bonds with high accuracy. Figure
\ref{fig:topoMethod} depicts the various intermediate stages in the
classification of fMSI formed within the graphene nano-capillary. Our
methodology is as follows.

First, all molecules connected by hydrogen bonds are identified using a
strict geometric criterion. Figure \ref{subfig:onlyHBN}(a) shows the HBN
for fMSI, with connections between oxygen molecules which are
hydrogen-bonded. In this particular treatment, no difference is made
between donor and acceptor molecules. We define an oxygen atom to be
hydrogen-bonded to a hydrogen atom if the angle between the O--O and O-H
vectors is less than 30 degrees, and if the distance between the
possible donor oxygen and acceptor hydrogen is less than \(2.42\)
\si{\angstrom} \cite{Chandra2003}.

Following this, we determine all possible primitive rings, from \(n=3\)
to \(n=10\). In this case, the number of primitive rings for \(n>4\) is
negligible. A distance cutoff of \(3.5\) \si{\angstrom} is used,
sufficient to encompass all first-neighbour shell molecules. However,
there are still several extraneous bonds, clearly visible in Figure
\ref{subfig:extraBonds}(c).

Next, we eliminate all primitive rings with connections between
molecules which are not hydrogen-bonded, using the previously determined
HBN information. By eliminating such non-hydrogen bonded rings, we
obtain precise connectivity information while still closely
approximating the HBN.

On eliminating extra bonds, the resultant ring network in Figure
\ref{subfig:awesomeAlgo}(d) almost perfectly coincides with the HBN in
Figure \ref{subfig:onlyHBN}(a). Our topological criterion correctly
classifies the monolayer ice as being primarily square ice, and entirely
eliminates the superfluous triangular phase.

\begin{figure}[H]
	\centering
	\includegraphics[width=0.8\textwidth,keepaspectratio]{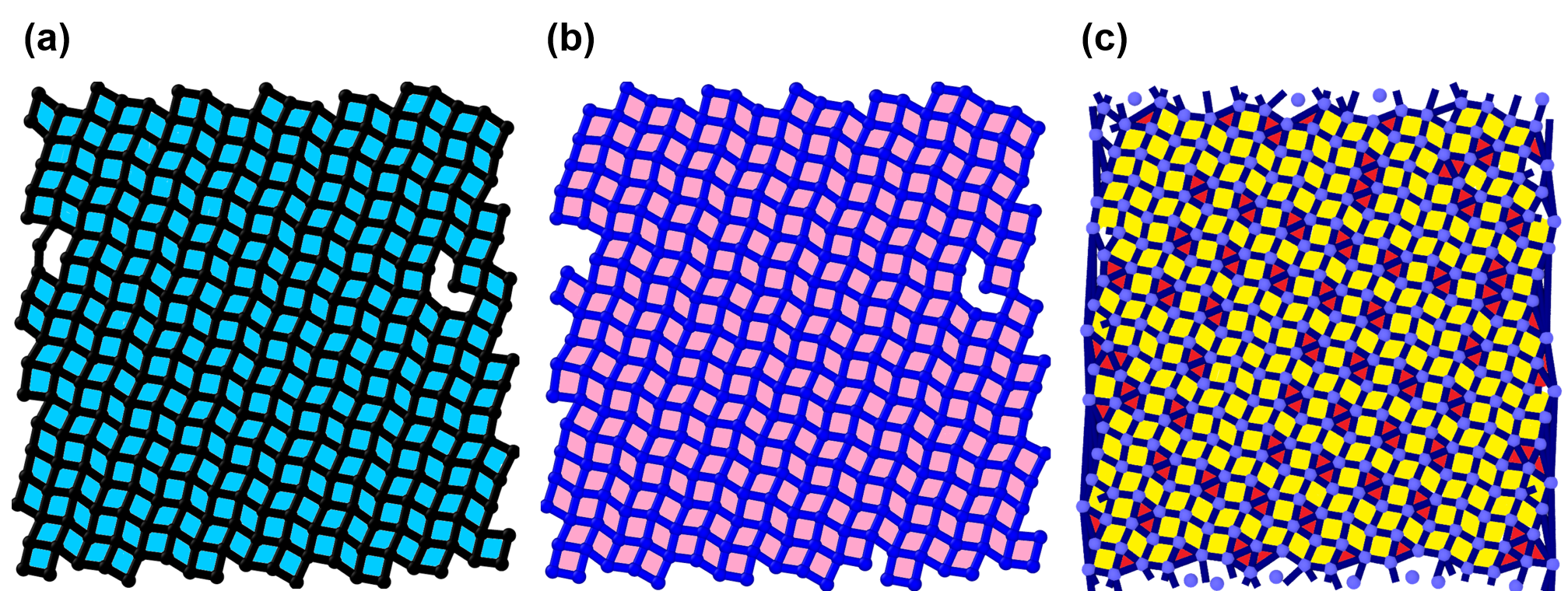}
	\caption{Comparison of the accuracy of Voronoi tessellation with our topological network criterion for identifying fMSI. fMSI is known to consist of 4-membered polygons. (a) The top view of the HBN, with
		4-membered polygons shaded in cyan. Identification schemes should closely approximate HBN connectivity. (b) The
		ring network generated by our topological network criterion, with 4-membered rings shaded in pink. (c) Neighbour bonds created by Voronoi tessellation. The bonds (in black) are generated by connecting neighbouring particles (blue) which share a Voronoi face. Polygons thus formed, with three and four edges, are coloured in red and yellow respectively. The Voronoi cells are 5-membered and 4-membered polygons, depending on the coordination number.}
	\label{fig:compareVoro}
\end{figure}

We have also tested our criterion by comparing its performance with that of Voronoi tessellation for
the same fMSI phase described above, visually represented in Figure \ref{fig:compareVoro}. The accuracy of
any identification scheme can be assessed by judging the extent to which it reproduces the connectivity of the HBN. Even after applying face area and edge length thresholds, Voronoi tessellation creates several extra connections (Figure \ref{fig:compareVoro}(c)). A perfectly accurate Voronoi diagram for this system would have had particles at the center of tetragonal Voronoi cells, corresponding to a coordination number of $4$ for each particle. In fact, the bond network generated by Voronoi tessellation in Figure \ref{fig:compareVoro}(c) closely resembles the primitive ring network of Figure \ref{subfig:noAlgo}(b), for whose generation the hydrogen bonds were not considered and only a fixed-distance cutoff was used.

The advantage of our topological network criterion for structural
determination is its unambiguous classification of the hydrogen-bonded
ring network in terms of $n$-membered primitive rings, even at ambient temperatures or strong confinement. It is also clearly
evident that, under such conditions, a direct application of a primitive ring analysis using a
fixed-distance cutoff introduces significant error. We observe that
there is especially an increase in the erroneous identification of the
triangular phase. Thus, keeping the non-hydrogen bonded extra connections (shown in
Figure \ref{subfig:noAlgo}(b)) causes the incorrect identification of
ice. We also note that the extra connections do not satisfy the
Bernal-Fowler ice rules \cite{Bernal1933}. The fMSI, which is known to
satisfy the ice-rules, visually indicates that the excess connections
are not concurrent with the ice rules, while the reduced ring network in Figure
\ref{subfig:awesomeAlgo}(d) does not, at least, appear to violate the ice
rules.

However, one caveat of eliminating rings is that, for example, one or
two 3-membered primitive rings may be removed, whose sides could have
formed a 4-membered ring which was not previously not considered to be
primitive according to the fixed-distance criterion. Such rings form a
negligible percentage of the whole, and are in the vicinity of point
defects or the edges. In this specific case, three sets of pairs of
triangles on the left edge and right edge were removed (visible when comparing Figures \ref{subfig:noAlgo}(b) and \ref{subfig:awesomeAlgo}(d)). These actually
form three 4-membered rings, which were also removed along with the extra
bonds.

Depending on the number of nodes (three or four) in each ring, we note
that the resulting topological network coincides with \(98-100 \%\) of
the HBN.

The ring network obtained from our topological network criterion may be used as-is for monolayer ice identification.
Using these hydrogen-bonded rings as a basis, we develop a prism identification algorithm for identifying $n$-gonal prism blocks.

\hypertarget{prism-identification-algorithm}{%
	\subsection{Prism Identification
		Algorithm}\label{prism-identification-algorithm}}

\begin{figure}[H]
	\centering
	\includegraphics[width=0.6\textwidth,keepaspectratio]{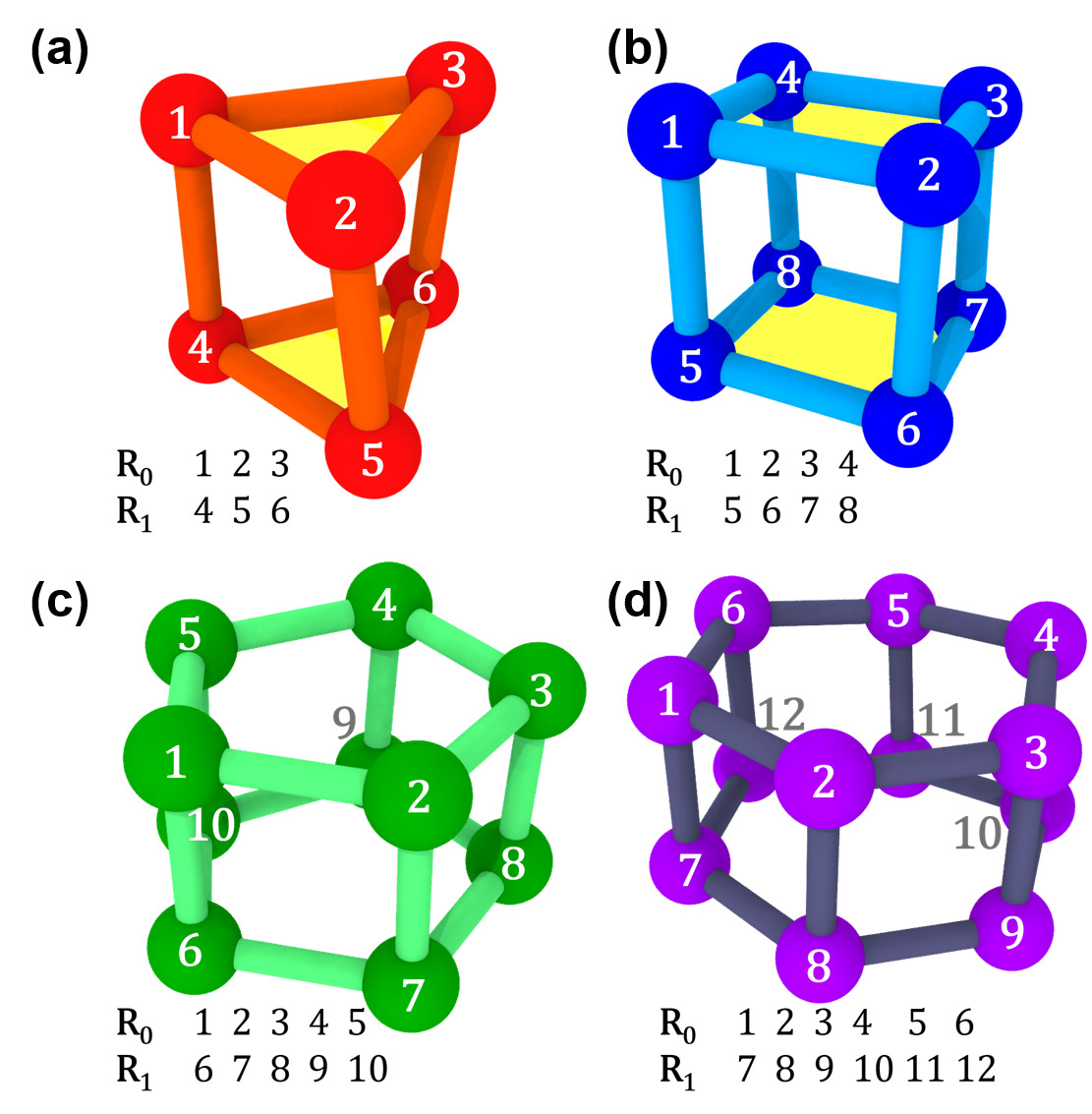}
	\caption{Topology of the building blocks of (a) triangular, (b) tetragonal, (c) pentagonal and (d) hexagonal prisms. The identification algorithm is the same for any $n$-sided prism. The basal faces of the prisms have been shaded in yellow for the triangular and tetragonal blocks. $R_0$ and $R_1$ are the basal planes of the prism. The basal planes have no nodes in common. Each node $m_k, k \in [1,n]$ in $R_0$ is hydrogen-bonded to a corresponding node $l_j, j \in [1,n]$ in $R_1$, where $k,j$ are integral indices. Every $m_k$ is only hydrogen-bonded to a single node $l_j$. In the case of tetragonal prisms, the basal faces and lateral faces are equivalent, but this does not affect the process of identification.}
	\label{fig:prismBlocks}
\end{figure}

The structural building block of an $n$-sided prism is an assortment of
two basal $n$-sided polygon planes, attached to each other by 4-membered
lateral rings, as shown in Figure \ref{fig:prismBlocks}. First,
hydrogen-bonded $n$-membered rings are obtained according to the procedure
outlined above in Section 2.1. These are subsequently used to identify
$n$-sided prism blocks.

\begin{figure}[H]
	\centering
	\includegraphics[width=\textwidth,keepaspectratio]{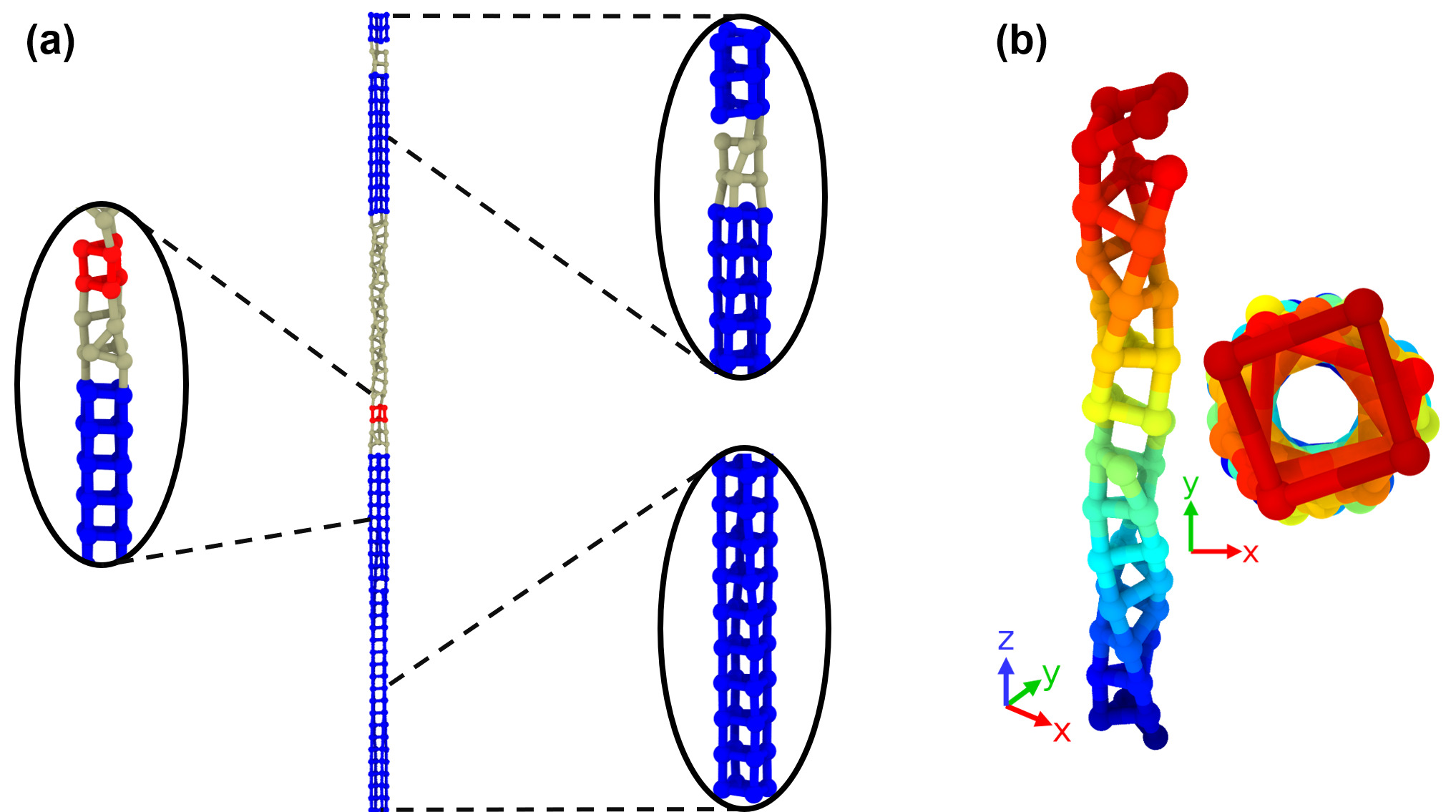}
	\caption{Structures formed within a zigzag $(13,0)$ SWNT, with explicit carbon atoms, at $280 K$. (a) The full transverse view of the nanoribbon, post-classification of prismatic building blocks. The insets show close-up perspective views of sections of the nanoribbon. The tetragonal prisms, triangular prism and unclassified rings are coloured in blue, red and grey respectively. The unclassified rings can denote either deformed prisms, helical order, or water molecules. The unclassified atoms in the upper right inset comprise a defective tetragonal prism. (b) Close-up perspective visual and top view of the unclassified atoms in the middle of the nanoribbon. They form a hollow helical structure exclusively composed of 4-membered rings. The atoms are coloured according to their heights.}
	\label{fig:testZigzag}
\end{figure}

We use the following recipe for identifying $n$-sided prisms. A pair of
potential basal faces \(R_0 = m_1, m_2, ..., m_n\) and
\(R_1 = l_1, l_2, ..., l_n\) is considered at a time. The 4-membered
lateral faces connecting each pair of basal planes in a prism unit may
be described solely by hydrogen bonds and do not need to be considered
separately. The methodology is the same for any integral value of \(n>2\).
Two candidate n-membered rings \(R_0\) and \(R_1\) are basal faces if
the following conditions are true.
\begin{enumerate}
	\def\labelenumi{\arabic{enumi}.}
	\item
	      \(R_0 \cap R_1 = \emptyset\). \(R_0\) and \(R_1\) should share no 	      common nodes. They are only connected by hydrogen bonds.
	\item
	      Every node \(m_k\) in $R_0$ must be a nearest neighbour of a
	      corresponding node \(l_j\) in \(R_1\), such that every node
	      in \(R_0\) is hydrogen-bonded to exactly one node in \(R_1\). For
	      example, a node \(l_j\) in \(R_1\) cannot be connected to both \(m_k\)
	      and \(m_{k+1}\) in \(R_0\).
\end{enumerate}
We apply our prism identification methodology to a zigzag \((13,0)\)
SWNT (single-walled nanotube), with \(R=13\) corresponding to a diameter of \(10.1\)
\si{\angstrom}. The structure of the ice nanoribbon at \(280 \ K\), with
a water occupancy of \(1.23 \ g/cm^3\), is shown in Figure
\ref{fig:testZigzag}. At this temperature, the nanoribbon is mostly
comprised of tetragonal prisms, one triangular prism, and
unclassified polygons in the middle. The rings which were classified as
belonging to no $n$-sided prisms (unclassified) may denote the water phase, deformed
prisms or a helical structure.

Specifically, the grey polygons in the
top right inset of Figure \ref{fig:testZigzag} are deformed tetragonal
prisms. We note also that the relatively twisted middle grey section of the
nanoribbon has a hollow tube section, which indicates that it may be an
ordered phase. It is also exclusively comprised of 4-membered polygons,
coiled in a helical structure. The methodology for handling unknown nanoribbons with disparate structure compositions is given by the algorithm listed below:

\begin{enumerate}
	\def\labelenumi{\arabic{enumi}.}
	\item
	      First, obtain the hydrogen-bonded $n$-sided primitive rings for the
	      system, such that $(n>2)$.
	\item
	      Apply the prism determination algorithm for every \(n\).
	\item
	      The unclassified structures remaining may be water, deformed prisms,
	      or part of an ordered ladder-like or helical structure. Liquid water
	      tends to form disordered rings of widely varying \(n\), filling up the
	      interior of the SWNT \cite{Takaiwa2007,Raju2018}. If the unclassified
	      structure does not fill the interior of the nanotube and exhibits a
	      certain degree of order, it may be part of a helical structure, or a
	      deformed prism, depending on the symmetry. A simple indication of
	      order is the number and type of primitive rings comprising
	      the unclassified sections. In the case of the nanoribbon in Figure
	      \ref{fig:testZigzag}, every unclassified molecule almost exclusively
	      participates in only 4-membered rings. We also show, in subsequent
	      sections, that our algorithm may be used to qualitatively describe the
	      abrupt symmetry change in freezing icy nanoribbons.
\end{enumerate}

\hypertarget{coverage-area-metric}{%
	\subsection{Coverage Area Metric}\label{coverage-area-metric}}

\begin{figure}[H]
	\centering
	\includegraphics[width=0.4\textwidth,keepaspectratio]{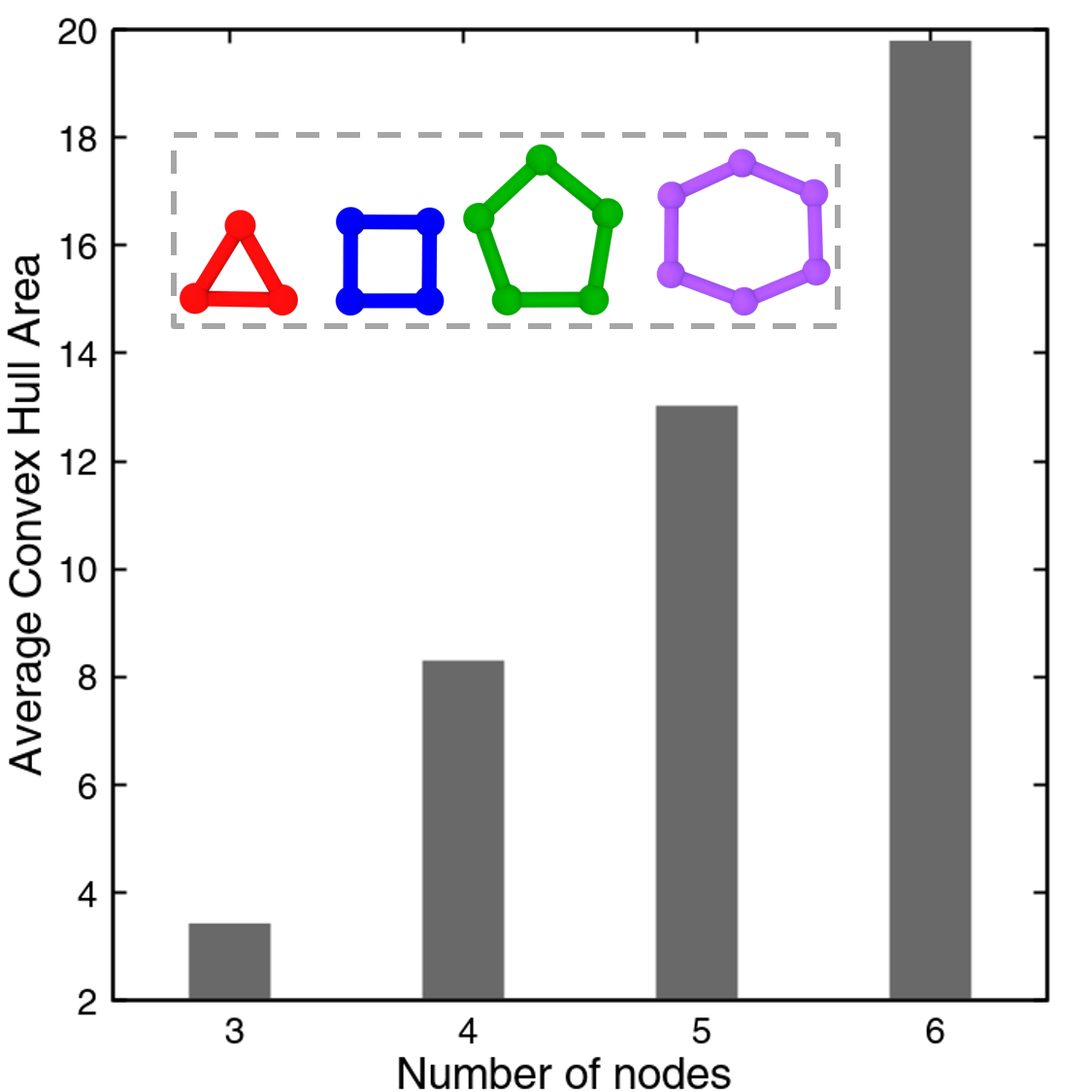}
	\caption{Average convex hull areas of $n$-membered rings observed in our systems. The convex hulls \cite{Barber96thequickhull} are formed from the vertices of the $n$-membered rings. 4-membered rings are approximately more than double the size of 3-membered rings, a conclusion which is supported by visual inspection of the HBN of fMSI within graphene nano-confinement. The insets show top views of typical $n$-membered rings observed in monolayer and nanoribbon ices.}
	\label{fig:singleAreaCompare}
\end{figure}

The number of rings has been used as a qualitative metric in previous
studies \cite{Chihaia2005,Salzmann2011,Zhang2017}. However, in our
systems, 3-membered rings in the HBN are smaller in size than 4-membered
rings, making the number of rings an unreliable descriptor (Figure
\ref{fig:singleAreaCompare}).

Intuitively, the area of each ring is a more qualitative metric for
monolayer systems. In our monolayer simulation systems, water layers are
formed in the \(XY\) plane. We define the \(XY\)-plane coverage area to
be the projected area of each ring onto the \(XY\) plane. We normalize
the coverage area by the theoretical maximum possible area of the ice,
which, in the case of monolayer water, is the area of the confining
sheet. The coverage area percentage is the percentage of the total area
of the constraining sheet that is comprised of the projected area of
$n$-membered rings, where \(n>2\).

\[Coverage \ Area \ Percentage_n = \frac{\sum_{i=1}^{i=N_n}{A_i}}{Area \ of \ sheet} \times100\]

where $A_i$ is the projected area of the $i^{th}$ $n$-membered ring, and
$N_n$ is the total number of $n$-membered rings.

The coverage area metric does not distinguish between small isolated
clusters of \(n\)-membered rings, and a contiguous connected cluster. It
is possible to apply additional constraints for connected rings of
certain types to identify ices like fMSI and pMSI (puckered Monolayer Square Ice), which are composed of
a single type of ring. We have observed that thermal
fluctuations momentarily fracture the ring network, which could disrupt
the connectivity of large swathes of ordered rings. For the monolayer
ices studied in this work, the coverage area percentage was found to
describe phase transitions and relative proportions of \(n\)-membered
rings accurately.

\hypertarget{occupied-volume-metric}{%
	\subsection{Occupied Volume Metric}\label{occupied-volume-metric}}

We formulate a qualitative volume-based metric to determine the relative proportions of
$n$-gonal prism structures within nanoribbons. Our prism determination
algorithm produces pairs of \(n\)-sided basal rings. We obtain the
volume enclosed by the prism block by calculating the volume of the
convex hull \cite{Barber96thequickhull} formed by the basal ring
vertices. For \(n \neq 4\), the basal rings are perpendicular to the
axis, but for \(n=4\), we only consider axial rings to prevent
overcounting.

For every \(n\), the total volume of each \(n\)-gonal prismatic phase is
obtained by summing over the convex hull volumes of the prism blocks. We
normalize this volume by the maximum possible volume, which is the occupied
volume if an \(n\)-sided prism were to fill the entire cylinder height. The
maximum possible length of each edge is \(3.5\) \si{\angstrom} from solvation
shell considerations. We have observed that the volume of each prism block for a particular \(n\) is relatively constant, thermal fluctuations
notwithstanding. Thus, for a particular prism phase, whose basal face has $n$ nodes, we can approximate the maximum possible filled volume by assuming that the SWNT is filled by a regular $n$-gonal prism of height $h$:

\[Maximum \ possible \ volume_n = Convex \ hull \ area \ of \ basal \ ring \times h\]

where \(h\) is the length of the SWNT.

Thus, we define the occupied volume percentage for an $n$-gonal prismatic phase as:

\[ Occupied \ Volume \ Percentage_n = \frac{\sum_{i=1}^{i=N_n}{V_i}}{Maximum \ possible \ volume_n} \times 100\]

where $V_i$ the convex hull volume of the $i^{th}$ $n$-gonal prism block, and $N_n$ is the total number of $n$-gonal prism blocks.

The maximum possible volume which we have used as a
normalizing factor varies for a particular \(n\). This is because the effective area
of the basal ring is different for every \(n\), with trigonal basal
rings occupying the least area. A consequence of this is that the sum of
the occupied volume of various composite \(n\)-gonal blocks
is always less than the total volume. We prove in subsequent sections, that this occupied volume percentage is an accurate qualitative
measure of phase transitions.

\hypertarget{simulation-details}{%
	\section{Simulation Details}\label{simulation-details}}

We have studied the following systems: monolayer ice systems with
explicit carbon atoms for simulating rigid (1) graphene layers, nanoribbon ices constrained by single-walled
nanotubes (2) with smooth featureless walls, and (3) with explicit
carbon atoms. The systems simulated are well-known in the literature and have thus been used for
validation and benchmarking.

\begin{figure}[H]
	\centering
	\includegraphics[width=0.5\textwidth,keepaspectratio]{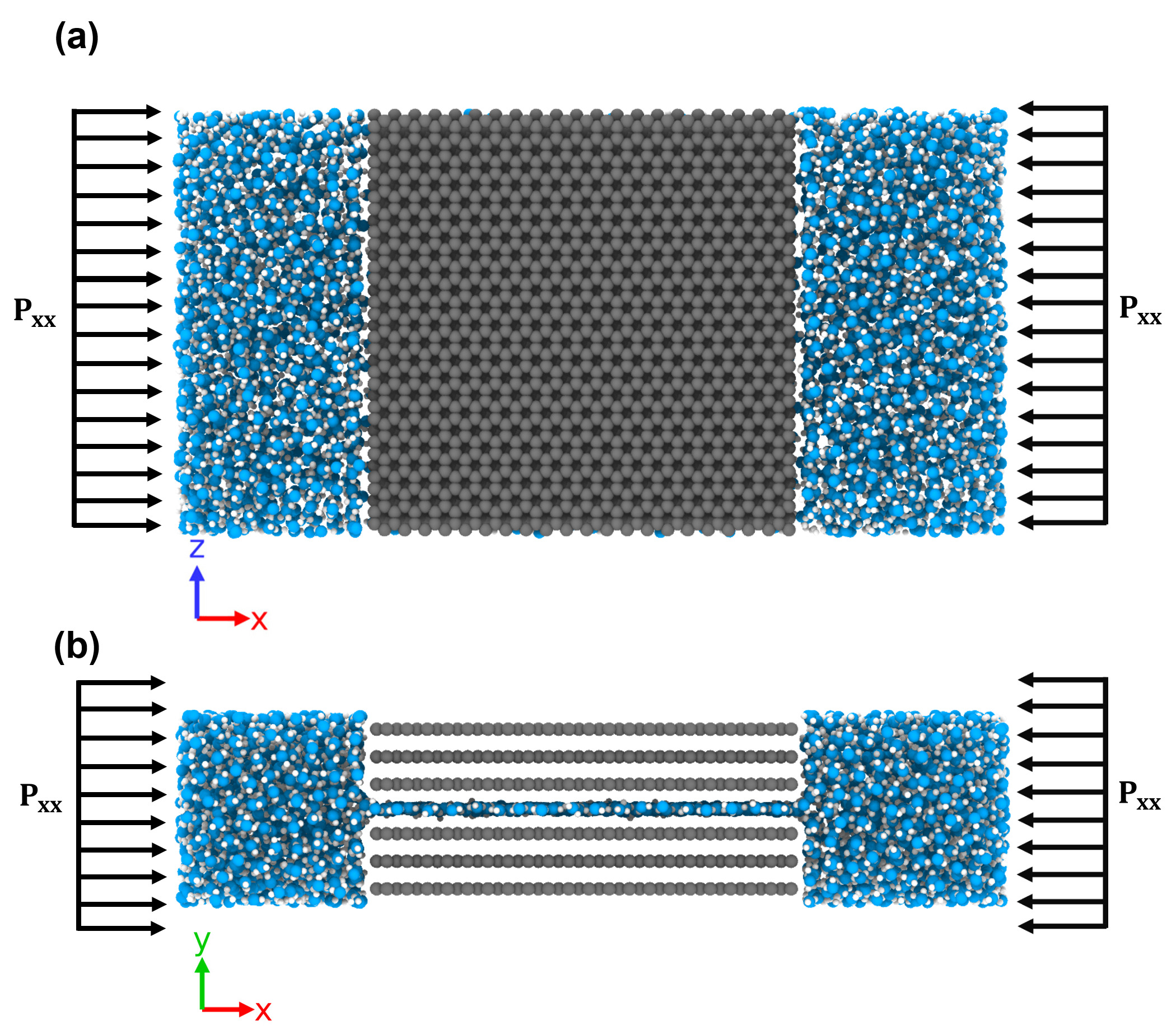}
	\caption{Representative views of the water monolayer constrained by graphene sheets. The graphene atoms are indicated in dark-grey; the oxygen atoms are light blue and the hydrogen atoms are white. (a) Top view of the simulation system, showing the XY plane. (b) Side view of the simulation system, showing the XZ plane. The center-to-center separation width between the parallel graphene layers is the gap separation width, $h$.}
	\label{fig:simSystem}
\end{figure}

For the monolayer ice graphene water system, the simulation setup is
similar to systems reported in the literature\cite{Yang2017}, shown in Figure
\ref{fig:simSystem}. A nanochannel was created by introducing a gap width \(h\)
between a pair of three graphene sheets. Two water reservoirs, containing
\(2000\) molecules, were placed on either side of the nanochannel thus formed.
The dimensions of the sheets were \(50.349\) \si{\angstrom} \(\times
50.349\) \si{\angstrom} in the \(X\) and \(Y\) dimensions, respectively. We used
center-to-center separation widths between the parallel diamond sheets of \(6\)
\si{\angstrom}, which is sufficient to permit the formation of a single layer of
water.

We have used LAMMPS \cite{Plimpton1995} to run equilibrium Molecular
Dynamics simulations. Lateral pressures were applied in the \(X\)
dimension. The MD simulations were performed in the isothermal-isobaric
\((NP_{xx}T)\) ensemble. The temperature \((T)\) and pressure
\((P_{xx})\) were controlled by the Nose-Hoover thermostat and barostat, respectively. A timestep of \(1 \ fs\) was used for the Velocity-Verlet
integration scheme. The particle-particle-particle mesh (PPPM) algorithm
was used for computing the long-range interactions.

TIP4P/2005 \cite{Abascal2005} was used to model the water molecules in
all the systems studied in this work. Among the commonly used rigid
water models used for simulating confined systems, TIP4P/2005 performs
reliably well compared to other popular rigid water models
\cite{Dix2018}. The sum of the interactions between the TIP4P/2005
molecules, and the external Lennard-Jones potential of interaction
between the water molecules and the confining sheet yield the total
potential of interaction.

In this work, we have used Lennard-Jones force-field parameters for the
interactions between the oxygen atoms of water and the carbon atoms of
graphene\cite{Werder2003}. These parameters have been extensively used
for graphite and water, accurately reproducing the contact angle of
water on graphite \cite{Yang2017,Zhu2015,Zhu2016,Zhu2017}.

For the SWCT simulations, the methodology of the previous literature was used \cite{Koga2001, Takaiwa2007},
and the featureless walls were simulated using the $9-3$ Lennard-Jones potential \cite{Kumar2005, Han2010}. The
explicit carbon atoms were modeled using the OPLS force field \cite{Jorgensen1996}.

The visuals in this work were created using OVITO \cite{Stukowski2009}.
Primitive rings were obtained using R.I.N.G.S \cite{Roux2010}. These
rings were further analysed according to our topological network
criterion.

\hypertarget{results-and-discussion}{%
	\section{Results and Discussion}\label{results-and-discussion}}

\hypertarget{freezing-of-an-ice-nanotube}{%
	\subsection{Freezing of an Ice
		Nanotube}\label{freezing-of-an-ice-nanotube}}

\begin{figure}[H]
	\centering
	\includegraphics[width=0.8\textwidth,keepaspectratio]{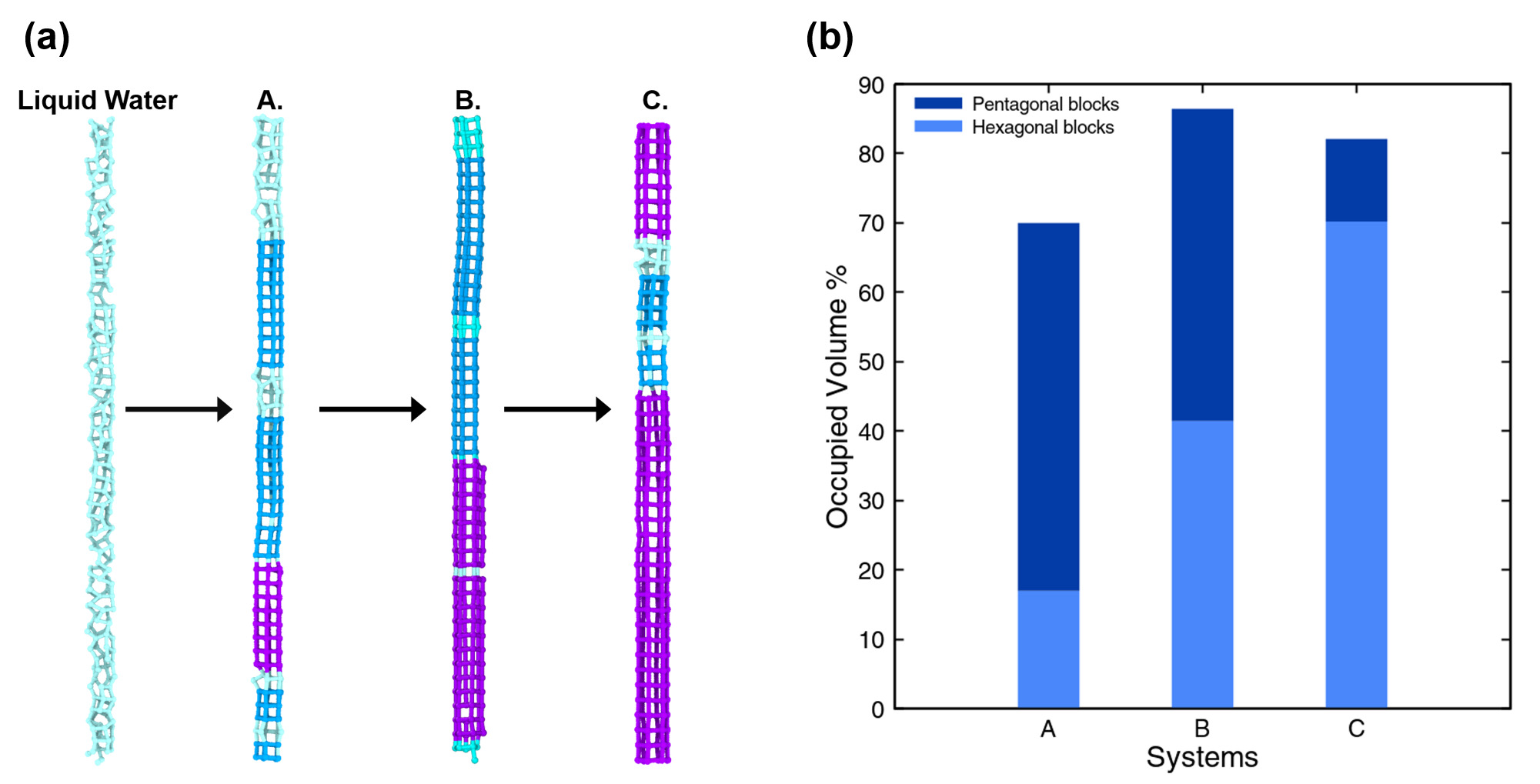}
	\caption{Pentagonal and hexagonal prismatic ice evolution in a $(13,0)$ featureless SWNT. (a) Snapshots of the water nanoribbon, as the temperature is lowered from $290 \ K$ to $240 \ K$. The particles and bonds are coloured in purple (hexagonal blocks), blue (pentagonal blocks) and light cyan (unclassified), as classified by our prism identification criterion. At $290 \ K$, the nanoribbon is completely water. Systems A and B correspond to the nanoribbon at $270 \ K$, just after abrupt freezing and after $20 \ ns$ respectively. System C shows the nanoribbon at $240 \ K$. (b) Occupied volume percentage for the pentagonal and hexagonal blocks, showing the change in relative proportions of the two ice types.}
	\label{fig:pentaHexaNT}
\end{figure}

We investigate the phase changes of water confined within a \((13,0)\)
featureless SWNT, corresponding to a diameter of \(10.1\)
\si{\angstrom}. When the temperature is lowered from \(320 \ K\), at a
constant axial pressure of \(10 \ MPa\), an abrupt change in state from
the liquid phase is observed at \(270 \ K\). The occupied volume
percentage rises suddenly from \(0 \%\) to \(\approx 53 \%\) and
\(\approx 17 \%\) for the pentagonal and hexagonal blocks respectively
(Figure \ref{fig:pentaHexaNT} (b)). No other \(n\)-gonal prism
structures are present.

The prism determination criterion is also able to capture solid-to-solid
phase transitions. Initially, at \(270 \ K\) (System A. in Figure
\ref{fig:pentaHexaNT}), a higher proportion of pentagonal prismatic ice
is observed. However, we observe a gradual change of the pentagonal
blocks to hexagonal blocks. After a \(20 \ ns\) simulation at
\(240 \ K\) the nanoribbon becomes almost exclusively comprised of
hexagonal prism blocks.

\hypertarget{monolayer-within-a-graphene-nanocapillary}{%
	\subsection{Monolayer Within a Graphene
		Nanocapillary}\label{monolayer-within-a-graphene-nanocapillary}}

Monolayer water exists in the form of fMSI when it is constrained
between two graphene surfaces \(6\) \si{\angstrom} apart at \(1 \ GPa\)
and \(300 \ K\) \cite{Zhu2015}. fMSI exhibits a first-order phase
transition when it is subjected to superheating at a constant low
lateral pressure (\(P_{zz} = 1 \ GPa\)) \cite{Zhu2017}.

\begin{figure}[H]
	\centering
	\includegraphics[scale=0.7]{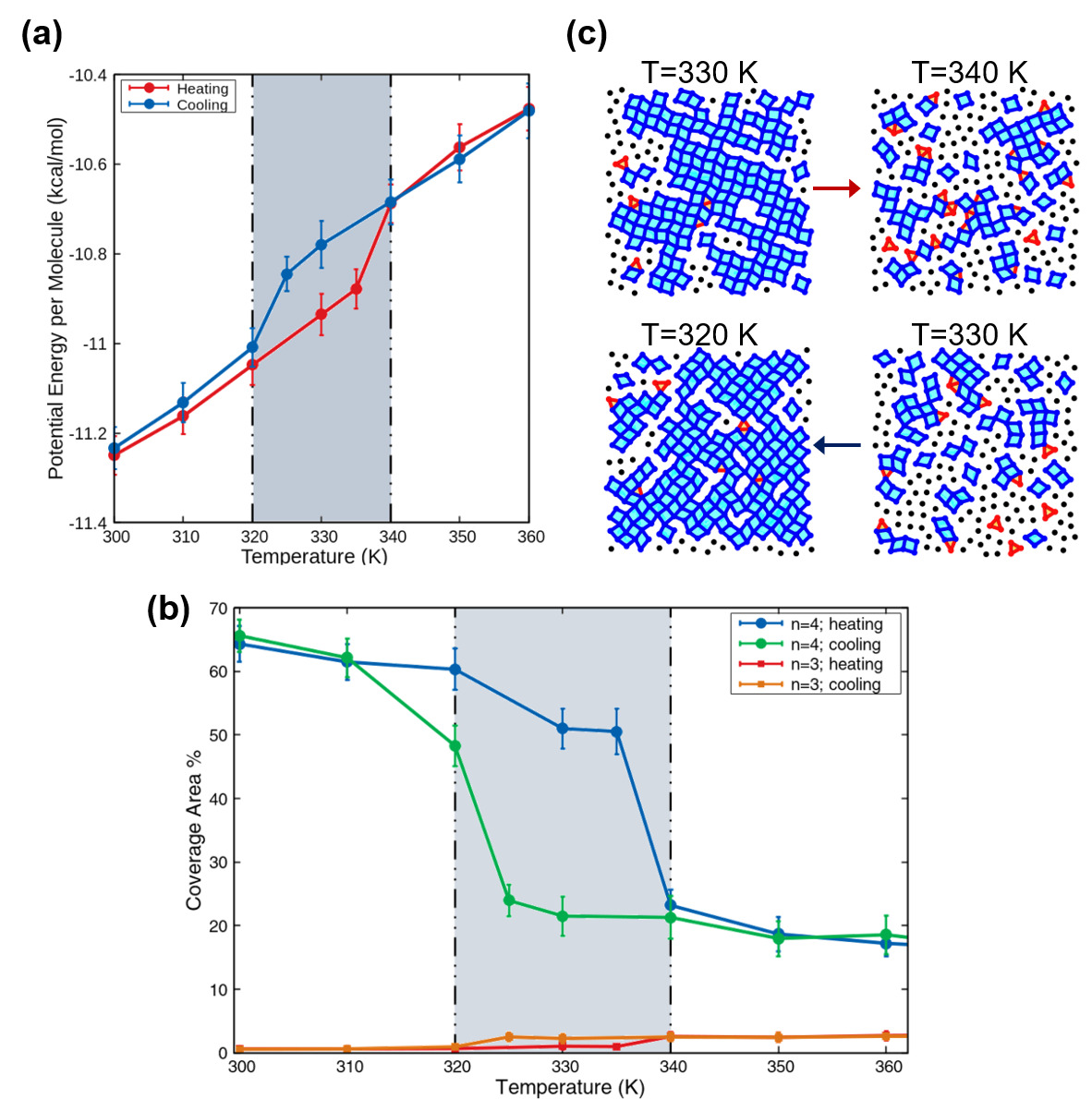}
	\caption{The first-order transitions observed in the heating and cooling of fMSI. (a) Potential energy per molecule change with temperature during the heating ( solid red line) and cooling (solid blue line) process. (b) The coverage area percentage plotted against temperature, for 4-membered and 3-membered rings. Here $n$ is the number of nodes in each type of ring. (c) Snapshots of the monolayer during the heating and cooling process, visually depicting the coverage area. The arrows are in the direction of increasing and decreasing temperature for the heating and cooling process respectively. 4-membered rings (blue) and 3-membered rings (red) are shaded in cyan and orange respectively. Particles which do not participate in the ring network are coloured in black.}
	\label{fig:grapheneLiqSol}
\end{figure}

We have performed heating and cooling MD simulations of fMSI, using a
simulation setup similar to those previously used in the literature
\cite{Zhu2015,Zhu2017,Yang2017}. Concomitant to the literature, we observe an abrupt increase in the
potential energy per molecule of \(\approx 0.2 \ kcal/mol\),
corresponding to the disruption of the ordered HBN (Figure
\ref{fig:grapheneLiqSol}(a)).

The plot of the coverage area percentage for 4-membered rings in Figure
\ref{fig:grapheneLiqSol}(b) shows a sharp decrease by \(\approx 25 \%\)
when the temperature is increased to \(340 \ K\) during the heating
process. The coverage area percentage of 3-membered rings also changes
during this phase transition. The disordered liquid network tends to
have a relatively higher proportion of 3-membered rings (visually
discernible in Figure \ref{fig:grapheneLiqSol}(c)). However, fMSI has
negligible proportions of 3-membered rings. When the monolayer is fMSI,
the coverage area percentage of the 3-membered rings remains constant at
\(\approx 0 \%\). The hysteresis loop of the 3-membered rings echoes that of the 4-membered rings and the potential energy per molecule.

\begin{figure}[H]
	\centering
	\includegraphics[scale=0.7]{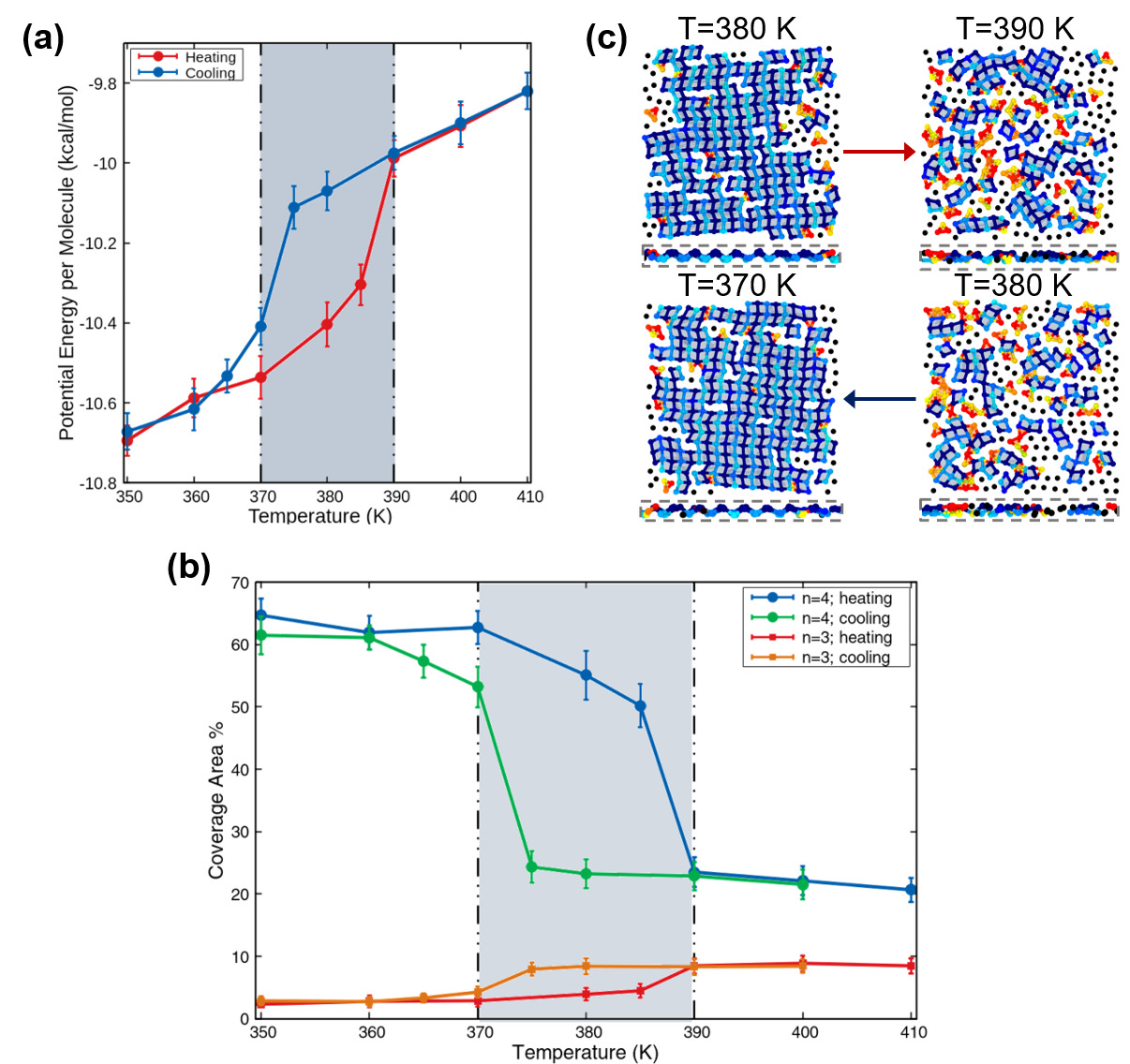}
	\caption{The phase transitions observed in the heating and cooling of pMSI at $2 \ GPa$. (a) Potential energy per molecule plotted against the temperature during the heating and cooling process. (b) The coverage area percentage change  with temperature, for 4-membered and 3-membered rings. (c) Snapshots of the puckered monolayer during the heating and cooling process, with the coverage area visually depicted by shaded-in $n$-membered rings. Arrows are in the directions of increasing and decreasing temperature for the heating and cooling process respectively. The insets show the puckered nature of the monolayer. Particles of the 4-membered rings and 3-membered rings are coloured in shades of blue and orange-red  respectively, according to the height. 4-membered and 3-membered rings are shaded in grey and orange respectively. Particles outside the ring network are coloured in black.}
	\label{fig:pmsiLiqSol}
\end{figure}

Figure \ref{fig:pmsiLiqSol}(a) shows the potential energy per molecule change during the heating and cooling of pMSI, at a constant lateral pressure of $2 \ GPa$. The coverage area percentage hysteresis loops mirror the hysteresis loop in the potential energy per molecule. The trends suggest that the phase transition is a first-order transition. The coverage area percentage change for the 3-membered rings (Figure \ref{fig:pmsiLiqSol}(b)) is more significant for pMSI than for fMSI (Figure \ref{fig:grapheneLiqSol}(b)).

Thus, the coverage area percentage is able to reproduce trends in the
potential energy per molecule, showing a correspondingly large
hysteresis loop.

\hypertarget{conclusions}{%
	\section{Conclusions}\label{conclusions}}

In this work we have presented a general and extensible family of topological network criteria for the
structural identification and analysis of quasi-one-dimensional and two-dimensional ices. The
primitive rings, identified using a fixed-distance cutoff, are screened to
obtain rings whose connections correspond to the HBN. An explicit accounting for the hydrogen bonds
in confined ordered structures has been found to be necessary. The effects of confinement may
force molecules to be clustered within a fixed distance cutoff corresponding to the first
nearest-neighbour shell, but are additionally constrained by the viability of hydrogen bonds. We show how the reduction of non-hydrogen bonded connections
is crucial to prevent mis-identification of fMSI. The primitive rings may be further
analysed for further identification, or used as-is.

We use the primitive rings so obtained
as a basis for an algorithm for identifying the building blocks of \(n\)-sided
prisms. Our prism identification algorithm classifies the \(n\)-membered prism
blocks which comprise single ice nanotubes (ICNs). Helical structures and
deformed prisms are not identified by this particular recipe, but can still
be distinguished from the liquid state by virtue of the number and type of the
component primitive rings. We show how a modified projected area, the coverage
area metric, may be used as an order parameter for investigating the phase
transitions in monolayer water. The coverage area percentage for every \(n\), is
a measure of the relative proportion of the $n$-membered rings. We have also
developed an analogous volume-based order parameter for ICN classification. In
the same spirit as that of the coverage area metric, we reinterpret the
topological features from a molecular simulation analysis perspective. The
occupied volume percentage is calculated using the convex hull volumes of
component prism blocks identified using the prism determination recipe. Both
these metrics are able to qualitatively describe various pressure-induced and
temperature-induced phase transitions.

The topological network criterion that forms the basis of these related
methodologies is flexible and easily customizable, requiring no a priori
assumptions beyond those for hydrogen bonds. We envisage that other
interconnected networks, including multicomponent structures and
solid-liquid interfaces may be probed using this general approach. The
relative speed of the criterion presented here, and the scope for
parallelization, make it possible to consider using it for on-the-fly
calculations during simulation runs, at least for small systems.

We anticipate that these sets of techniques may have a wider
applicability for other confined structures and geometries, especially
those with mixed structures of unknown relative composition.

\hypertarget{acknowledgements}{%
	\subsection*{Acknowledgements}\label{acknowledgements}}

This work was supported by the Department of Science and Technology,
Govt of India. Computational resources were provided by the HPC cluster
of the Computer Center (CC), Indian Institute of Technology Kanpur.

\printbibliography

\end{document}